\documentclass[prl,letter,twocolumn,showpacs,lengthcheck,superscriptaddress]{revtex4}
\usepackage{graphicx}
\usepackage{amsmath,amssymb}
\usepackage{bm}
\usepackage{ifthen}

\newcommand{\tr}{\operatorname{Tr}}

\newcommand{\ket}[1]{| #1 \rangle}

\begin{document}
\title{Theory of excitonic 
spectra and entanglement engineering in dot molecules}
\author{Gabriel Bester}
\affiliation{National Renewable Energy Laboratory, 
Golden CO 80401}
\author{J. Shumway}
\affiliation{National Renewable Energy Laboratory, 
Golden CO 80401}
\affiliation{Department of Physics and Astronomy,
Arizona State University, Tempe AZ 85287-1504}
\author{Alex Zunger}
\affiliation{National Renewable Energy Laboratory, Golden CO 80401}
\date{\today}

\begin{abstract}
We present results of
correlated pseudopotential calculations of an exciton in a pair of vertically 
stacked InGaAs/GaAs dots.
Competing effects of strain, geometry, and band mixing lead to many unexpected
features missing in contemporary models. 
The first four excitonic states are all optically active at
small interdot separation, due
to the broken symmetry of the single-particle states.
We quantify the degree of entanglement of the 
exciton wavefunctions and show its sensitivity to
interdot separation. We suggest ways to spectroscopically 
identify and maximize the entanglement of exciton states.
\end{abstract}


\maketitle
 
The small size of semiconductor quantum dots \cite{Bimberg99}
drives speculations that they
may provide a physical representation of a quantum bit (qubit) that 
supports a superposition of ``0'' and ``1'' 
\cite{loss98,zinardi98,imamoglu99,zinardi99,biolatti00,troiani03}. 
Some proposed representations of qubits include 
electron spin states \cite{loss98} and
the presence or absence of an electron, hole, or electron-hole pair 
(exciton).\cite{bayer01,korkusinski01,korkusinski02,bayer02b}
Registers of qbits might be realized in {\em coupled quantum dots},
self-assembled by strain-driven islanding of InGaAs on a 
GaAs substrate \cite{shtrichman02}. In one possibility,
an electron represents qubit $A$ and a hole represents 
qubit $B$, while the qubit states are the occupation of
either the top ($T$) or bottom ($B$) 
dot \cite{bayer01,korkusinski01,korkusinski02,bayer02b}.
This quantum register must store entangled states. Predicting
entanglement requires a
theory of the electronic structure of the dot-molecule, including
single-particle and correlation effects. Most 
modeling of dot-molecules has been done in 
single-band effective-mass approximation
\cite{fonseca98,yannouleas99,koskinen03,troiani02,troiani03,fujisawa02}.
For two equivalent dots, this treatment leads to 
single-particle electron and hole orbitals 
forming bonding and antibonding combinations:
\begin{alignat}{2}
\label{eq:sp_states}
\ket{\phi_{\text{h}}^{\text{b}}} &= \textstyle \frac{1}{\sqrt{2}}
 (\ket{h_{\text{T}}} + \ket{h_{\text{B}}}),&\;\; 
\ket{\phi_{\text{h}}^{\text{a}}} &= \textstyle \frac{1}{\sqrt{2}} 
 (\ket{h_{\text{T}}} - \ket{h_{\text{B}}}) \nonumber \\
\ket{\phi_{\text{e}}^{\text{b}}} &= \textstyle \frac{1}{\sqrt{2}} 
 (\ket{e_{\text{T}}} + \ket{e_{\text{B}}}),&
\ket{\phi_{\text{e}}^{\text{a}}} &= \textstyle \frac{1}{\sqrt{2}} 
 (\ket{e_{\text{T}}} - \ket{e_{\text{B}}})
\end{alignat}
where $e_{\text{T}}$ ($e_{\text{B}}$) represents an electron
in the top (bottom) dot and $h_{\text{T}}$ ($h_{\text{B}}$) represents a hole in the top 
(bottom) dot. 
In this picture, as in an H$_2^+$ molecule,
the single-particle {\em bonding} state energy $E_{\text{e}}^{\text{b}}(d)$
decreases as the interdot distance $d$ decreases
while the bonding hole state energy $E_{\text{h}}^{\text{b}}(d)$ increases.
{\em Simple direct products of single-particle states}, e.g.
$\ket{\phi_{\text{e}}^{\text{b}}\phi_{\text{h}}^{\text{b}}} 
= \frac{1}{2}[\ket{e_{\text{T}}h_{\text{T}}} + \ket{e_{\text{B}}h_{\text{B}}}
 + \ket{e_{\text{T}}h_{\text{B}}} + \ket{e_{\text{B}}h_{\text{T}}}]$, 
are {\em unentangled}.
In contrast, the desirable {\em maximally entangled Bell}
states superpose either exciton or dissociated states, but not both:
\begin{alignat}{2}
\label{eq:fully_entangled}
\ket{a} &= \textstyle \frac{1}{\sqrt{2}} (\ket{e_{\text{B}}h_{\text{B}}} + \ket{e_{\text{T}}h_{\text{T}}});&\;
&\text{exciton, bonding,} \nonumber\\
\ket{b} &= \textstyle \frac{1}{\sqrt{2}} (\ket{e_{\text{B}}h_{\text{T}}} + \ket{e_{\text{T}}h_{\text{B}}});&\;
&\text{dissociated, bonding,} \nonumber\\
\ket{c} &= \textstyle \frac{1}{\sqrt{2}} (\ket{e_{\text{B}}h_{\text{T}}} - \ket{e_{\text{T}}h_{\text{B}}});&\;
&\text{dissociated, antibonding,}\nonumber\\ 
\ket{d} &= \textstyle \frac{1}{\sqrt{2}} (\ket{e_{\text{B}}h_{\text{B}}} - \ket{e_{\text{T}}h_{\text{T}}});&\;
&\text{exciton, antibonding.}
\end{alignat}

Recent optical experiments on vertically-stacked
double dots \cite{bayer01,bayer02b} 
claimed to show entangled excitonic states, but
the evidence for entanglement is indirect and based on a symmetric  model 
underlying Eq.~(\ref{eq:sp_states}). 
Unlike simple symmetric molecules like H$_2^+$,
double-dot ``molecules'' of stacked InGaAs dots are made
of $\sim$$10^5$ atoms and have
complicated interactions such as alloy fluctuations, strain, multi-band
(e.g. light-heavy hole), inter-valley ($\Gamma$-$X$), and 
spin-orbit couplings not included in the symmetric molecular case. 
To properly simulate these double dots, we have performed detailed atomistic
pseudopotential calculations, including correlation, on a realistic dot molecule. In this
Letter we report on new insights into the exciton state: 
all states are  optically active at short distances,
entanglement is small except at a critical dot separation $d_c$
at which the low energy exciton is darkened, yielding
a spectroscopic signature of entanglement.

\begin{figure} 
\includegraphics[width=0.90\linewidth]{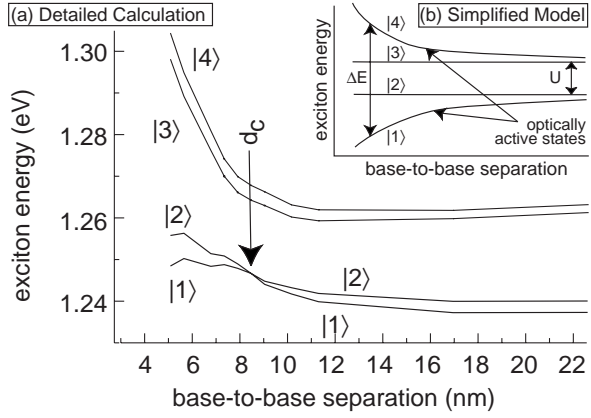}
\caption{Exciton energies as a function of the interdot separation for
(a) our pseudopotential many-body results.\label{fig:exciton}
(b) a model calculation $\Delta E=\sqrt{(4t)^2+U^2}$.}
\end{figure}

To understand previous theoretical treatments
of excitons in a dot molecule and to set a reference to which our atomistic
results 
will be compared, we describe a generic two-site tight-binding 
Hamiltonian in the basis of  products of 
electron and hole single-particle states 
$\ket{e_{\text{T}} h_{\text{T}}}$, $\ket{e_{\text{B}} h_{\text{T}}}$,
 $\ket{e_{\text{T}} h_{\text{B}}}$, $\ket{e_{\text{B}} h_{\text{B}}}$:
\begin{equation}\label{eq:hamiltonian}
H=\begin{pmatrix}
E_{\text{eh}}^{\text{TT}} & t_{\text{e}} & t_{\text{h}} & 0 \\
t_{\text{e}} & E_{\text{eh}}^{\text{BT}} & 0 & t_{\text{h}}\\
t_{\text{h}} & 0& E_{\text{eh}}^{\text{TB}} & t_{\text{e}}\\
0& t_{\text{h}} & t_{\text{e}} & E_{\text{eh}}^{\text{BB}}
\end{pmatrix}
\end{equation}
\begin{alignat*}{2}
E_{\text{eh}}^{\text{TT}} &=\varepsilon_{\text{e}}^{\text{T}} 
- \varepsilon_{\text{h}}^{\text{T}} + U_{\text{eh}}^{\text{TT}},&\quad
E_{\text{eh}}^{\text{BT}} &=\varepsilon_{\text{e}}^{\text{B}} 
- \varepsilon_{\text{h}}^{\text{T}} + U_{\text{eh}}^{\text{BT}} \\
E_{\text{eh}}^{\text{TB}} &= \varepsilon_{\text{e}}^{\text{T}}
- \varepsilon_{\text{h}}^{\text{B}} + U_{\text{eh}}^{\text{TB}},&\quad
E_{\text{eh}}^{\text{BB}} &= \varepsilon_{\text{e}}^{\text{B}} 
- \varepsilon_{\text{h}}^{\text{B}} + U_{\text{eh}}^{\text{BB}}+\Delta E
\end{alignat*}
Here $\{\varepsilon_{\text{e}}^{\text{T}}, \varepsilon_{\text{e}}^{\text{B}},
 \varepsilon_{\text{h}}^{\text{T}}, \varepsilon_{\text{h}}^{\text{B}}\}$ are the electron 
and hole on-site energies, $\{t_{\text{e}},t_{\text{h}}\}$ are the tunneling matrix elements, and 
$\{U_{\text{eh}}^{\text{TT}},U_{\text{eh}}^{\text{TB}},
U_{\text{eh}}^{\text{BT}},U_{\text{eh}}^{\text{BB}}\}$
are the electron-hole Coulomb matrix elements.
The extra parameter $\Delta E$ that will be used later in
entropy discussion; initially we set $\Delta E=0$.
A simplification, followed in Refs.~[\onlinecite{bayer01,korkusinski02,bayer02b}], 
is to set: $t_{\text{e}}=t_{\text{h}}=t$, 
$\varepsilon_{\text{e}}^{\text{T}} = \varepsilon_{\text{e}}^{\text{B}}$ and
$\varepsilon_{\text{h}}^{\text{T}} = \varepsilon_{\text{h}}^{\text{B}}$, 
with intra-dot Coulomb energies 
$U_{\text{eh}}^{\text{TT}} = U_{\text{eh}}^{\text{BB}}=U$, and 
neglecting inter-dot terms
$U_{\text{eh}}^{\text{TB}}$ and $U_{\text{eh}}^{\text{BT}}$. 
With this simplification, the Hamiltonian of Eq.~(\ref{eq:hamiltonian}) 
yields, in increasing order of energy, the four $e$-$h$ states:
\begin{equation}
\label{eq:eigenvectors_model2}
\begin{split}
\ket{1} = & \textstyle (\ket{a} - \gamma_1\ket{b})/\sqrt{1+\gamma_1^2}, \\
\ket{2} = & \textstyle \ket{c},\quad \ket{3} =  \ket{d},\\
\ket{4} = & \textstyle (\ket{a} - \gamma_2\ket{b})/ \sqrt{1+\gamma_2^2},\\
\gamma_{1,2} =& \left[U\pm\sqrt{(4t)^2+ U^2}\right]/4t.
\end{split}
\end{equation}
We see that in the simplified model states $\ket{2}$ and $\ket{3}$ are fully
entangled pure Bell states [viz. Eq.~(\ref{eq:fully_entangled})] that are spatially 
antisymmetric (anti-bonding) and therefore optically dark.
In contrast, states $\ket{1}$ and $\ket{4}$ are not fully entangled
and have some symmetric (bonding) character making them optically allowed.
Assuming that the tunneling integral $t(d)$ decays
with interdot separation $d$, the simple model gives the level order shown in
Fig.~\ref{fig:exciton}b). In the simple model, the exciton
$\ket{1}$ shifts to the red as $d$ decreases, and the separation of the 
two bright states $\ket{1}$ and $\ket{4}$ increases as
$\Delta E = \sqrt{(4t)^2 + U^2}$. Furthermore, the order of
the levels is $\ket{1}$=bright $\rightarrow \ket{2}$=dark
$\rightarrow \ket{3}$=dark$\rightarrow \ket{4}$=bright. 
This is in apparent 
agreement with experiments that show the same qualitative behavior 
\cite{bayer01,korkusinski02,bayer02b},
spurring hope that the theoretically predicted high 
degree of entanglement in this system could be 
experimentally realized to the benefit of quantum computing.
\begin{figure}
\includegraphics[width=0.90\linewidth]{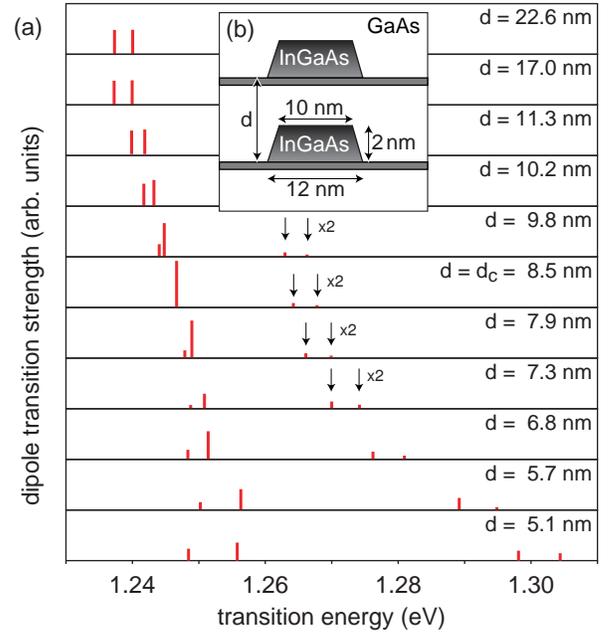}
\caption{(a) Emission spectra in a pair of vertically stacked
InGaAs/GaAs dots. (b) Dot geometry, including a two monolayer
(0.56 nm) InGaAs wetting layer and graded 
composition profile.\label{fig:spectra}}
\end{figure}

There are reasons to doubt the simple homonuclear 
diatomic-like analogue of dot
molecules of Eq.~(\ref{eq:sp_states}) and Fig.~\ref{fig:exciton}(b).
Actual self-assembled quantum dots contain $\sim$$10^5$
atoms, while the 
dots themselves are strained by the host matrix and subjected to random 
alloy fluctuations. 
Thus, a ``molecule'' made of two dots does
not necessarily behaves like homonuclear H$_2$, but
could behave like a heteronuclear molecule (e.g.~HF) since strain and alloy
fluctuations distinguish the dots, $\varepsilon^{\text{T}}\ne\varepsilon^{\text{B}}$.
Furthermore, the electronic properties of
such dots cannot \cite{wang99c}  be accurately modeled by simple single-band
effective-mass approaches: coupling between a large number of bands 
alters electron and hole localization, changing the Coulomb matrix elements.
Finally, the assumption of equal tunneling for electron and hole, 
$t_{\text{e}}=t_{\text{h}}$, is  questionable given the
large mass ratio, $m_{\text{e}}/m_{\text{hh}}\approx 1/6$,
of electrons and heavy holes in the GaAs barrier between
the dots. In fact, we see below that band mixing even changes the {\em sign} of $t_{\text{h}}$.
Thus, a more complete theoretical treatment is warranted.

\begin{figure}
\includegraphics[width=0.9\linewidth]{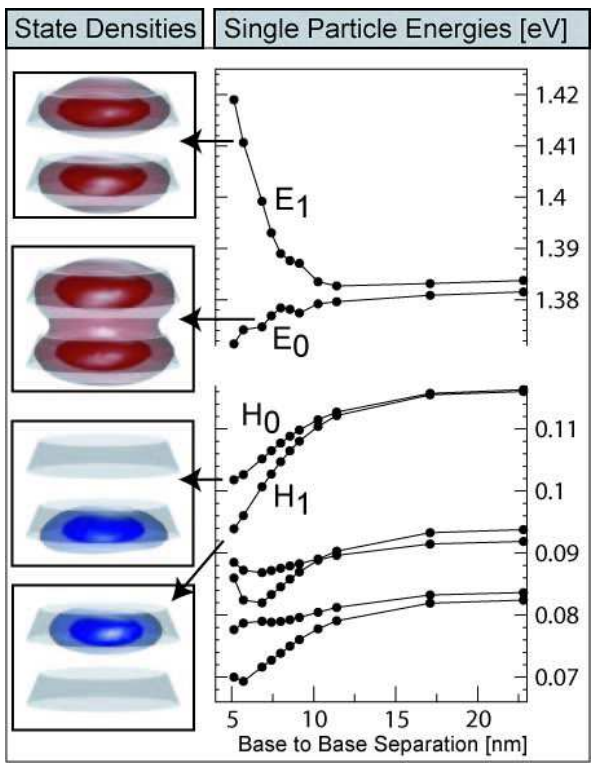}
\caption{Single-particle energies as a function of interdot separation and 
single-particle state densities from our pseudopotential calculations. 
The two translucent isosurfaces enclose 75\% and 40\% of the total 
state densities. The physical dot dimension are shown in grey, with
base-to-base separation d=5.1 nm.}
\label{fig:single_particle}
\end{figure}

We simulate the InGaAs/GaAs
dot molecule at a range of inter-dot spacings, using a computational approach
that successfully describes single InGaAs/GaAs 
dots \cite{williamson01}. 
Specifically, we describe the single-particle properties with
an atomistic empirical pseudopotential method, with
the wavefunctions expanded in
a set of Bloch states of the constituent materials over many bands and wave
vectors \cite{wang9599b}.
The theory includes multi-band and multi-valley
coupling, spin-orbit interaction, and anisotropic strain effects. 
To calculate correlated $e$-$h$ states, we include excitonic effects in a low-order
configuration interaction expansion, as in Ref.~\onlinecite{franceschetti99},
calculating all Coulomb and exchange integrals explicitly from the
single-particle wavefunctions. The dot geometry has been chosen 
to resemble the experimental system studied by Bayer {\em
et al} \cite{bayer01}.
As shown in  Fig.~\ref{fig:spectra}(b), the dots are 12 nm 
$\times$ 2 nm truncated-cone-shaped, 
with a linear composition gradient varying from In$_{0.5}$Ga$_{0.5}$As
at their bases to pure InAs at their tops. 
Fig.~\ref{fig:single_particle} shows the calculated 
single-particle energies and wavefunctions. 
We plot our calculated correlated $e$-$h$ energies, Fig.~\ref{fig:exciton}(a)
and corresponding absorption spectra, Fig.~\ref{fig:spectra}.

\begin{figure}
\includegraphics[width=0.92\linewidth]{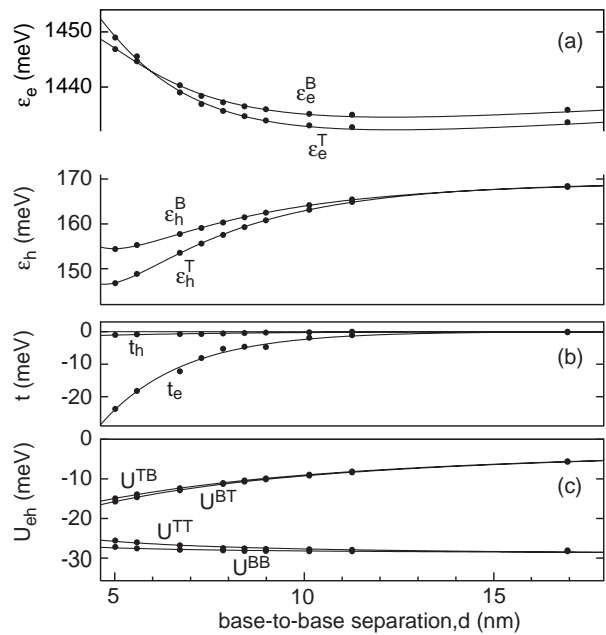}
\caption{Effective parameters for the two-site Hamiltonian,
Eq.~(\ref{eq:hamiltonian}), distilled from our pseudopotential calculations.}
 \label{fig:hamiltonian}
\end{figure}

By projecting our numerically calculated correlated $e$-$h$  energies vs.~$d$ onto a
generalized form of $H$, Eq.~(\ref{eq:hamiltonian}), 
we have determined the effective distance-dependent Hamiltonian parameters, 
shown in Fig.~\ref{fig:hamiltonian}. These represent realistic values for the
simplified model parameters contemplated in Ref.~\onlinecite{loss98}.
Inspection of the parameters from Fig.~\ref{fig:hamiltonian} 
and the comparison within Fig.~\ref{fig:exciton} show that our results differ in  crucial ways 
 from the simple assumptions made in 
Refs.~\onlinecite{bayer01,korkusinski02,bayer02b}.
We next discuss the salient features of these differences and their physical
implications.

\paragraph{(i) The energies of exciton $\ket{1}$ and $\ket{2}$ blue-shift
as $d$ decreases, in contrast with the red-shift expected from the 
simple model.}
Two effects are responsible for this surprising 
blue-shift. (a) From a {\em single-particle view}, Fig.~\ref{fig:single_particle} shows 
that as the interdot separation decreases, the energy of 
both hole states $h_0$ and 
$h_1$ move to lower values, while the molecular bonding-antibonding
picture of Eq.~(\ref{eq:sp_states}) would predict that the bonding 
$h_0$ level will moves to higher energies. The
downward shift of the single-particle hole level $h_0$ with decreasing $d$
contributes to the upward shift of the lowest excitons $\ket{1}$ and 
$\ket{2}$ observed 
in Fig.~\ref{fig:exciton}(a). The reason that the single-particle
states $h_0$, $h_1$ move to lower energies as $d$ decreases is 
their {\em symmetry
broken} character: $h_0$ is localized on $B$ and 
$h_1$ on $T$, as seen in the densities,
Fig.~\ref{fig:single_particle}. This localization is reflected in  the small
tunneling matrix element for holes $t_{\text{h}}$, 
in Fig.~\ref{fig:hamiltonian}, and is due to the
heavier hole mass and the strong strain-induced potential barrier 
\cite{sheng02b} between the dots.
In contrast, the light-mass electrons have a large tunneling matrix
element
$t_{\text{e}}$ and 
follow the bonding-antibonding picture of Eq.~(\ref{eq:sp_states}), as
evidenced by the calculated density,
Fig.~\ref{fig:single_particle}, exhibiting 
delocalization on both dots.
(b) From an {\em interacting-particle view}, the blue shift
of $\ket{1}$ and $\ket{2}$ with decreasing $d$ is caused by the decrease
in  the Coulomb 
elements $U_{\text{eh}}^{\text{TT}}$  and $U_{\text{eh}}^{\text{BB}}$ 
with reduced interdot separation
(shown in Fig.~\ref{fig:hamiltonian}(c)), due to delocalization of the exciton
on both dots. 

\paragraph{(ii) At large $d>10 $ {\em nm} the order of 
excitons $\ket{1}$, $\ket{2}$, $\ket{3}$ and
$\ket{4}$ is bright, bright, dark and dark, 
in contrast with the simple model
predicting the order bright, dark, dark, bright.}
The large $d$ behavior of our pseudopotential calculations 
can be understood  in the tight-binding language:
differences in on-site energies 
$(\varepsilon_{\text{e}}^{\text{T}}-\varepsilon_{\text{h}}^{\text{T}})-
(\varepsilon_{\text{e}}^{\text{B}}-\varepsilon_{\text{h}}^{\text{B}})$
are greater than hopping elements $t_{\text{e}}$ and $t_{\text{h}}$,
(Fig.~\ref{fig:hamiltonian}).
With these assumptions, the exciton states,
in increasing order of energy, 
are given by $\ket{1} = \ket{e_{\text{T}} h_{\text{T}}}$, 
$\ket{2} = \ket{e_{\text{B}} h_{\text{B}}}$,
$\ket{3} = \ket{e_{\text{T}} h_{\text{B}}}$, $\ket{4} = \ket{e_{\text{B}} h_{\text{T}}}$. 
States $\ket{1}$ and 
$\ket{2}$ are  bright since they are symmetric and have large $e$-$h$ overlap.
In contrast, states $\ket{3}$ and $\ket{4}$ are dark since they are not symmetric and have low $e$-$h$ overlap.
These four eigenstates are obviously {\it not}
entangled, while the simple model predicts full entanglement. 

\paragraph{(iii) Exciton $\ket{1}$ and $\ket{2}$ anticross at the critical
distance $d_c$ at which point $\ket{1}$ becomes dark.  
However, all excitons are bright at $d<d_c$, in 
contradiction with the simple model.}
This can be  understood as follows. 
At $d_c$ ($\simeq$8.5 nm for our specific case) the basis states 
$\ket{e_{\text{T}}h_{\text{T}}}$ and $\ket{e_{\text{B}}h_{\text{B}}}$ 
 are nearly degenerate. 
{Now hopping elements $t_e$ and $t_h$ will split this near degeneracy
into} symmetric and antisymmetric combinations 
$\ket{2} 
=\ket{a}$ and
$\ket{1} 
=\ket{b}$, respectively.
Whether the ground state is symmetric or antisymmetric is decided by the
respective signs of $t_e$ and $t_h$. If they have 
opposite signs, the symmetric $e$-$h$ state has lower energy. 
This is the case in the single band effective mass 
approximation \cite{troiani02} 
where the single-particle hole {\em and} electron states have 
pure S-envelope-character creating 
$V_{ss\sigma}$ bonds \cite{Harrison80}  
with {\em positive} $t_h$ and negative $t_e$.  However, in a realistic case, 
the presence  of inter-band mixing \cite{wang99c} (S-P as well
as heavy and light hole) leads to a  $V_{pp\sigma}$-like hole bond, with 
$t_h<0$. In fact, the confining potential for holes
attracts the light-hole P-like 
component of the hole states to the bonding region in-between the dots.
If $t_e$ and $t_h$ have the same sign (viz, our 
Fig.~\ref{fig:hamiltonian}b), 
the antisymmetric state (dark) is below the symmetric (bright) state .

\begin{figure}
\includegraphics[width=0.92\linewidth]{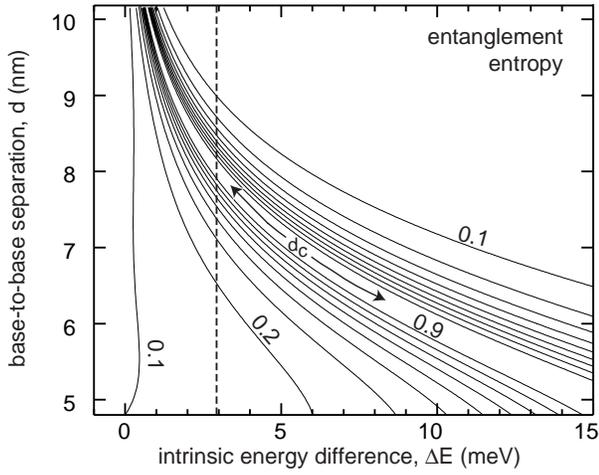}
\caption{Degree of entanglement after Eq.~(\ref{eq:entropy}). 
The dashed line indicates the value 
$\Delta E=\varepsilon_{\text{e}}^{\text{T}} 
- \varepsilon_{\text{e}}^{\text{B}}$=2.9 meV  from our simulations. 
\label{fig:entropy}}
\end{figure}
\paragraph{(iv) The degree of entanglement reaches a maximum at a critical
interdot separation $d_c$:}
Since $\ket{1}$ and $\ket{2}$ approach Bell-states $\ket{a}$
and $\ket{d}$ at $d=d_c$, we expect high entanglement.
We quantify the degree of entanglement
directly from our pseudopotential $e$-$h$ density 
matrix, using the von Neumann entropy of entanglement \cite{bennett96},
\begin{equation}\label{eq:entropy}
S=-\tr \rho_{\text{e}} \log_2 \rho_{\text{e}} = -\tr \rho_{\text{h}} \log_2 \rho_{\text{h}},
\end{equation}
where $\rho_{\text{e}}$ and $\rho_{\text{h}}$ are the reduced density matrices
of the electron and hole, respectively. 
For exciton $\ket{1}$ we find $S=0\%$
for $d>10$ nm and a pronounced peak at $d_c$ with $S=80\%$.
$d_c$ is determined by a balance between 
interdot strain coupling
and intrinsic dot energy differences (alloy fluctuations, in our calculations).
To generalize our results to a {\em class} of dots we allow  in
Eq. (\ref{eq:hamiltonian})  a generic fluctuation $\Delta E$.
We have calculated $S(d,\Delta E$), as shown in Fig.~\ref{fig:entropy},
using our fitted model Hamiltonian, Eq.~(\ref{eq:hamiltonian}) 
and Fig.~\ref{fig:hamiltonian}. This shows how various degrees of entanglement
can be engineered.
We note that the specific case $d_c=8.5$ nm arises from our intrinsic
dot energy separation $\Delta E=\varepsilon_{\text{e}}^{\text{T}} 
- \varepsilon_{\text{e}}^{\text{B}}$=2.9 meV.

Extracting the entanglement from this exciton state for use in quantum
computing 
may require the separation of the electron and hole while maintaining
phase coherency. This experimental challenge might be accomplished  
by driving the particles to nearby dot molecules using an in-plane
electric field.

In conclusion, we find that the entanglement entropy reaches a maximum 
value (of 80\% in
our case) at a critical interdot separation and decays abruptly 
to zero at smaller and larger separations. We suggest that the distance $d_c$
can be  
identified using photoluminescence spectra, by noting two closely
spaced exciton peaks with a {\em darker} lower energy peak. 

\acknowledgments
Work supported by the DOE SC-BES-DMS Grant No. DE-AC36-99GO10337.

\bibliographystyle{apsrev}

\end{document}